\documentclass[11pt]{article}

\usepackage[preprint]{acl}

\usepackage{times}
\usepackage{latexsym}

\usepackage[T1]{fontenc}

\usepackage[utf8]{inputenc}

\usepackage{microtype}

\usepackage{inconsolata}

\usepackage{graphicx}
\usepackage{booktabs}
\usepackage{multirow}
\usepackage{adjustbox}
\usepackage[table]{xcolor}
\usepackage[most]{tcolorbox}
\usepackage{pifont}

\newcommand{\cmark}{\ding{51}} 
\newcommand{\xmark}{\ding{55}} 
\tcbuselibrary{listings,breakable}
\title{Audio-Oscar: A Multi-Agent System for Complex Audio Scene Generation, Orchestration, and Refinement}

\author{
 \textbf{Yifan Duan\textsuperscript{1,2,\thanks{Equal contribution.}}},
 \textbf{Qixiang Xu\textsuperscript{1,\footnotemark[1]}},
 \textbf{Hengtao Wu\textsuperscript{1,\footnotemark[1]}},
 \textbf{Zhanxun Liu\textsuperscript{1,2,3}},
 \\
 \textbf{Wenhao Guan\textsuperscript{2,4}},
 \textbf{Junxi Liu\textsuperscript{1}},
 \textbf{Ziyang Ma\textsuperscript{1,2}},
 \textbf{Kelu Xu\textsuperscript{5}},
 \textbf{Xie Chen\textsuperscript{1,2,\thanks{Corresponding author.}} },
\\
 \textsuperscript{1}MoE Key Lab of Artificial Intelligence, X-LANCE Lab, Shanghai Jiao Tong University,\\
 \textsuperscript{2}Shanghai Innovation Institute,
 \textsuperscript{3}Shanghai AI Laboratory,
 \textsuperscript{4}Xiamen University,\\
 \textsuperscript{5}State Key Laboratory of Complex \& Critical Software Environment, China
\\
\texttt{ziye\_hitsz@163.com, chenxie95@sjtu.edu.cn}
}

\begin{document}
\maketitle
\begin{abstract}
In recent years, audio generation has made significant progress in tasks such as text-to-speech (TTS), text-to-audio (TTA) and text-to-music (TTM). However, generating long-form and controllable audio from complex audio scene descriptions remains a significant challenge, as such scenes often require coordinated speech, sound effects, music, songs, temporal structure, and post-production. In this work, we introduce \textbf{Audio-Oscar}, a multi-agent framework for generating audio from complex descriptions. Audio-Oscar coordinates a set of specialist agents, each responsible for a different aspect of the audio scene, including character modeling and voice design, speech generation, fine-grained timeline planning, model selection, non-speech generation, and audio post-production. Audio-Oscar further incorporates feedback-driven refinement. In addition, to address the lack of suitable benchmarks for evaluating audio generation from complex audio scene descriptions, we construct \textbf{ASG-Bench}, an Audio Scene Generation Benchmark containing both scene descriptions paired with reference audio and text-only scene descriptions. Each scene is annotated with target audio events and temporal statements to evaluate whether the generated audio faithfully realizes the required scene content and temporal structure. Experimental results show that Audio-Oscar can effectively generate audio that matches complex scene descriptions. Project samples are available at \url{https://audiooscar.github.io/}. Our code is available at \url{https://github.com/ziye26/Audio-Oscar}.

\end{abstract}

\section{Introduction}
Audio is an important component of multimedia content and plays an important role in scenarios such as videos and podcasts. In recent years, with the rapid development of various generative models, audio generation has made significant progress in tasks like text-to-speech (TTS), text-to-audio (TTA) and text-to-music (TTM)~\cite{chen2025f5,hu2026qwen3,liu2024audioldm,evans2025stable,yuan2026yue}. The models can now effectively generate high-quality speech, environmental sounds and music content~\cite{du2025cosyvoice,li2025meanaudio,gong2026ace}. However, in complex audio scenarios, audio is often composed of a combination of multiple sound elements, such as speech, background music, ambient sounds and event sound effects. Existing audio generation models often focus on single task generation and are unable to meet the demands of generating complex audio.

\begin{figure}[t]
  \centering
  \includegraphics[width=\linewidth]{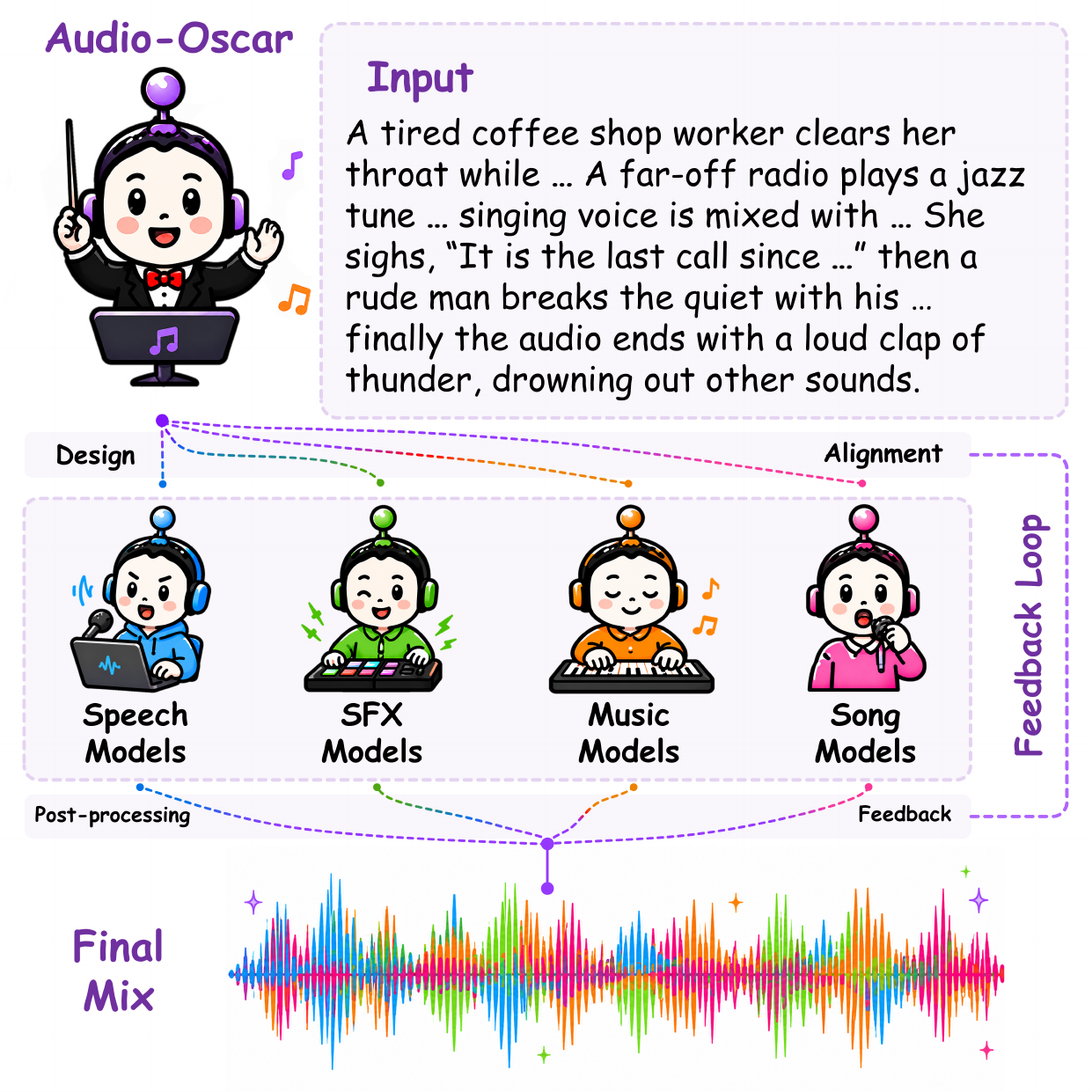}
  \caption{Overview of Audio-Oscar. Given an audio scene description as input, it coordinates multiple agents and employs different models for generation, composition, and refinement to produce the final audio output.}
  \label{fig:intro}
\end{figure}

Increasingly, recent works have attempted to use a unified model to cover different types of audio generation, including speech, music, and environmental sounds, and have begun to show preliminary capabilities to generate complex mixed audio~\cite{liu2024audioldm,yang2023uniaudio,vyas2023audioboxunifiedaudiogeneration,ai2025ming}. Meanwhile, recent speech generation research has also started to extend toward fuller and richer acoustic scene modeling. For example, Any2Speech~\cite{song2026borderless} further introduces the modeling of acoustic contexts such as background sounds and environmental ambience, enabling the generated outputs to present more complete acoustic scenes. 

LLM-based agents have developed rapidly and have been widely applied to solve various complex tasks~\cite{yao2023react,yang2024swe,lian2025mmmem}. Recent research has begun to explore using agents to address audio generation tasks. Early works such as AudioGPT~\cite{huang2024audiogpt} utilized large language models (LLMs) as a universal interactive interface to resolve different types of audio generation tasks. WavJourney~\cite{liu2025wavjourney} leveraged LLMs for compositional audio creation. Meanwhile, a wide range of agent-based methods have been proposed and have achieved great success in tasks such as podcast generation, audiobook production, and video dubbing~\cite{rong2025audiogenie,xiao2025podagent,rong2026dopamine,ren2026audirectorselfreflectiveclosedloopframework}. However, these methods often rely on predefined voice or voice library, face challenges in accurately planning timelines for complex audio scene descriptions, and often lack the capability to autonomously perform audio post-processing. 

Directly generating final audio from complex audio scene descriptions remains a significant challenge. To address this challenge, we propose \textbf{Audio-Oscar}, a framework that \textbf{O}rchestrates multiple \textbf{S}pecialized agents for \textbf{C}omplex \textbf{A}udio scene generation and \textbf{R}efinement. Audio-Oscar transforms complex scene descriptions into a structured generation process through role and voice modeling, fine-grained timeline planning, and expert model orchestration for different audio types. It further incorporates an audio post-production agent and feedback-driven refinement to improve the naturalness and overall quality of the generated audio. In addition, to address the lack of suitable benchmarks for evaluating audio generation from complex scene descriptions, we introduce the \textbf{A}udio \textbf{S}cene \textbf{G}eneration \textbf{Bench}mark (\textbf{ASG-Bench}). ASG-Bench includes audio scene descriptions with reference audio as well as text-only scene descriptions. Each scene is annotated with target audio events and temporal statements to evaluate whether the generated audio faithfully realizes the required scene content and temporal structure. Our experiments show that Audio-Oscar can effectively generate audio that reflects complex scene descriptions.

Our contributions are as follows:
\begin{itemize}
    \item We propose \textbf{Audio-Oscar}, a multi-agent collaborative system that uses specialized agents for scene orchestration, audio synthesis, post-production, and mixing to generate audio based solely on complex audio scene text descriptions.
    \item We propose \textbf{ASG-Bench}, a benchmark designed to evaluate the ability of models and systems to generate extended audio from complex audio scene descriptions.
    \item Our experimental results demonstrate that Audio-Oscar can effectively generate complex, long-form audio that faithfully follows the given descriptions.
\end{itemize}

\section{Related Works}
\subsection{Audio Generation Models}
Recent advances in TTS, TTA, and TTM have achieved remarkable success. Recent progress in TTS has achieved breakthroughs in zero-shot expressivity, streaming latency, and long-form synthesis~\cite{chen2025f5, peng2025vibevoice, du2025cosyvoice, hu2026qwen3}. TTA models have demonstrated great capabilities in generating high-quality audio that is well aligned with textual prompts, while also improving inference efficiency~\cite{liu2024audioldm, li2025meanaudio, evans2025stable, cheng2025mmaudio}. Meanwhile, TTM models have shown strong capabilities in generating high-quality music with coherent structures and improved controllability over lyrics, instrumentation, and style~\cite{yuan2026yue, gong2026ace, copet2023simple, liu2025songgen}. However, these specialized models often remain restricted to single task generation. Building on these advances, recent audio-generation architectures have begun to synthesize speech, music, and sound effects within a unified modeling framework~\cite{qiang2026unisonate, ai2025ming}. However, when applied to more complex audio scenarios, existing models still face challenges in maintaining long-form compositional coherence and modeling persistent yet dynamically evolving environmental sounds.

To expand modality coverage and interactivity, recent work like Any2Speech~\cite{song2026borderless} extends speech synthesis toward borderless long-audio generation, using structured semantic controls to model acoustic environments and scene-level context. Furthermore, joint video-audio generation has achieved remarkable cross-modal synchronization, enabling the generation of high-quality and complex audio~\cite{chern2026speed, seedance2026seedance}. Despite this progress, these models are often restricted to generating only short clips, requiring storyboard switching or external techniques for longer durations. Moreover, they produce inherently mixed audio without clear track separation, severely complicating subsequent post-production.

\subsection{Audio Generation Agents}

\begin{figure*}[t]
    \centering
    \includegraphics[width=0.98\textwidth]{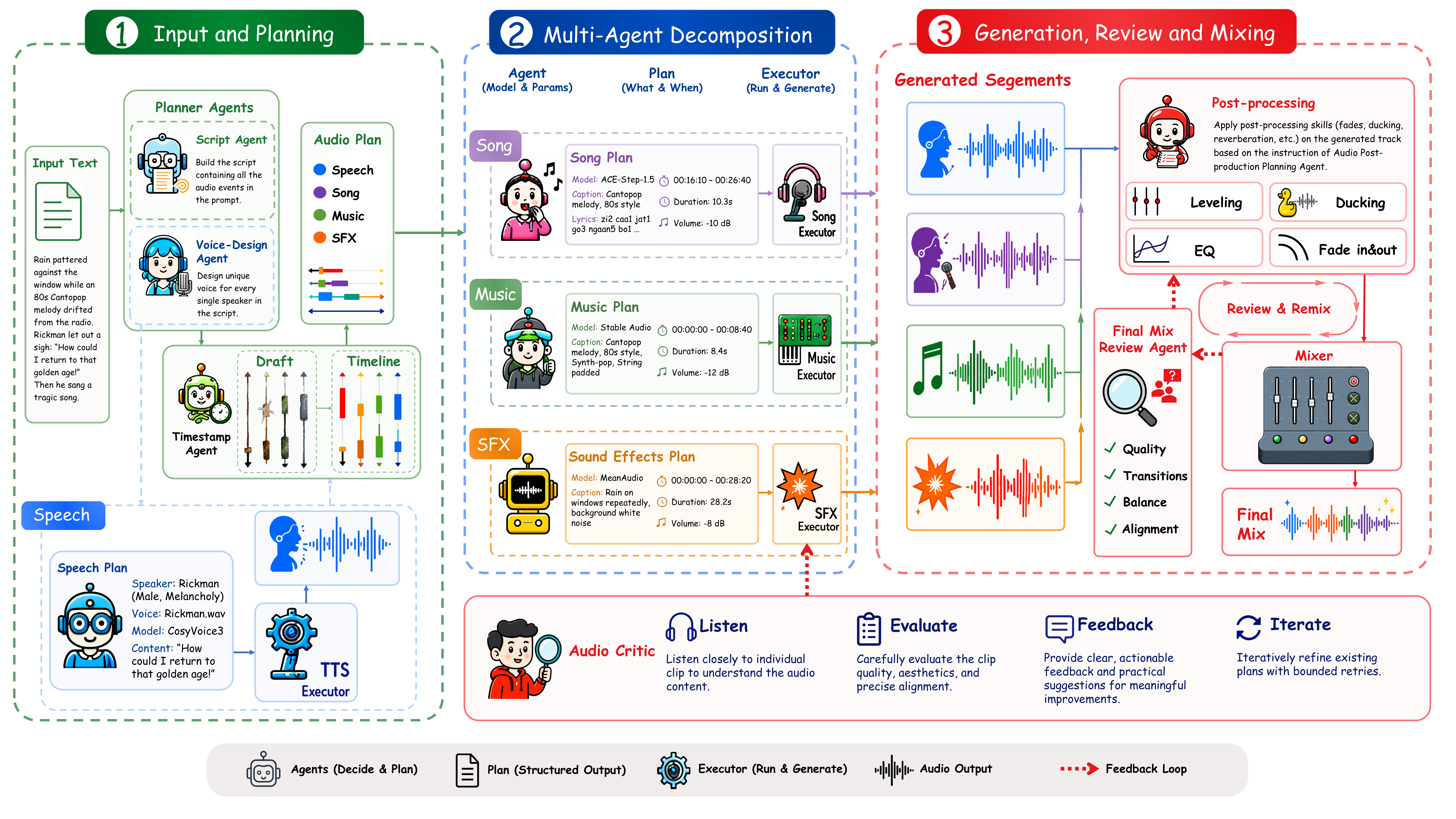}
    \caption{Overview of Audio-Oscar. Given a complex audio scene description, Audio-Oscar decomposes the generation process into multiple coordinated stages, including speaking role extraction and voice design, speech planning and generation, timeline construction and refinement, non-speech audio planning and generation, edit-aware mixing, and feedback-driven refinement.}
    \label{fig:system-overview}
\end{figure*}

Driven by the rapid advancements in Large Language Models (LLMs), autonomous agents have experienced exponential development and widespread adoption across numerous fields~\cite{yao2023react,yang2024swe,lian2025mmmem}. In the acoustic domain, this emergence of LLM-based agents has opened a new paradigm for compositional audio generation by orchestrating foundational models for complex tasks~\cite{huang2024audiogpt}. Notably, WavJourney~\cite{liu2025wavjourney} pioneered this space by decomposing complex audio scenes into executable scripts, demonstrating that LLMs can logically plan combinations of speech, music, and sound effects. Subsequent multi-agent systems have introduced role-specific collaboration to tackle long-form podcast generation~\cite{xiao2025podagent}, developed a video-to-audio agentic system~\cite{rong2025audiogenie}, achieved word-level temporal alignment for emotional audiobooks~\cite{rong2026dopamine} and integrated closed-loop self-reflection mechanisms for interactive audio storytelling~\cite{ren2026audirectorselfreflectiveclosedloopframework}.

However, existing agentic systems still struggle to directly generate audio from complex audio scene descriptions and often lack post-production capabilities for audio refinement. These limitations motivate Audio-Oscar, a multi-agent framework designed to generate complex audio from audio scene descriptions.

\section{Overview}

Audio-Oscar is a multi-agent system for complex audio scene generation. As shown in Figure~\ref{fig:system-overview}, it decomposes an input scene description into a structured production workflow rather than generating the entire scene in a single pass. The system first normalizes the input, builds speaking role and voice profiles, and generates speech audio at the utterance level. The measured speech durations then serve as temporal anchors for drafting and refining a full audio timeline that jointly schedules speech, sound effects, music, and songs.

Given the refined timeline, Audio-Oscar routes different audio segments to specialized generation modules, applies critic-guided repair for low-quality sound effects when needed, and augments the timeline with segment-level post-production instructions. The generated segments are finally mixed into a complete scene-level audio output, with a final audit-and-remix step for local post-production refinement. This design enables our system to effectively generate complex audio scenes.

\section{Method}

In this section, we provide a detailed overview of the generation process in Audio-Oscar. More details about the agents involved in this process are provided in Appendix~\ref{apd:audiooscar}.

\subsection{Role Modeling and Voice Design}
Given an input audio scene description $x$, Audio-Oscar first identifies the speaking roles and constructs role-level voice profiles. The Speaker Role Profiling Agent extracts the set of speakers
$$
\mathcal{R}=\{r_i\}_{i=1}^{M}
$$
from the input description and produces a profile for each speaker:
\[
v_i=\mathcal{A}_{\mathrm{role}}(r_i,x),
\]
where $v_i$ contains character names, character description, and voice characteristics. These profiles provide speaker-level conditions for downstream speech generation.

Audio-Oscar then invokes a voice-design model to synthesize a reference voice for each role:
\[
c_i=\mathcal{A}_{\mathrm{voice}}(v_i),
\]
where $c_i$ denotes the generated reference audio. This stage allows the system to maintain consistent speaker identities across multi-speaker scenes.

\subsection{Speech Planning and Generation}
As the actual duration of generated speech is difficult to predict precisely during planning, Audio-Oscar first plans and generates the speech at the utterance level to enable more reliable temporal control. The speech event planning agent extracts explicit dialogue or spoken narration from the input and constructs a set of speech elements
\[
\mathcal{E}_{\mathrm{sp}}=\{e_i^{\mathrm{sp}}\}_{i=1}^{N_{\mathrm{sp}}}.
\]
Each speech element is represented as
\[
e_i^{\mathrm{sp}}=(s_i, q_i, r_i)
\]
where $s_i$ is a stable segment identifier, $q_i$ is the utterance text, and $r_i$ is the speaker. 
For each speech element, the speech generation planner selects a suitable TTS model and generation parameters according to the utterance text, speaker, input scene, and model capabilities. The selected TTS model then generates the speech.
After synthesis, each speech result is augmented with its generated duration:
\[
\hat{e}_i^{\mathrm{sp}}=(s_i, q_i, r_i, d_i)
\]
where $d_i$ is the measured duration of the synthesized speech. These speech results are later passed to the audio planning stage, which places them on the global timeline by choosing start times and deriving end times.

\subsection{Audio Timeline Planning and Refinement}
Given the generated speech results, Audio-Oscar constructs a full audio timeline that combines speech and non-speech events. Following~\citet{rong2025audiogenie}, we categorize non-speech audio into three types: sound effects (SFX), music, and songs, and define type-specific fields with reference to AudioGenie. 

The audio timeline drafting agent takes the input scene description and the synthesized speech results $\hat{\mathcal{E}}_{\mathrm{sp}}=\{\hat{e}_i^{\mathrm{sp}}\}_{i=1}^{N_{\mathrm{sp}}}$ as constraints, and produces an initial timeline:
\[
\mathcal{T}^{(0)}=\mathcal{A}_{\mathrm{audio}}(x,\hat{\mathcal{E}}_{\mathrm{sp}})=\{z_i\}_{i=1}^{N}.
\]
Each element $z_i$ specifies an audio event in the timeline, including its audio type, start time, end time, duration, and other information such as volume and description.

Speech results are treated as protected elements. The agent preserves all content and use their actual durations. It only chooses their start times and the end times are derived from the measured durations.
For each speech segment, the planner decides whether forced alignment is needed to construct a more precise timeline.

After the draft timeline is produced, a timestamp refinement agent repairs and refines the schedule:
\[
\mathcal{T}=\mathcal{A}_{\mathrm{time}}(x,\mathcal{T}^{(0)},\hat{\mathcal{E}}_{\mathrm{sp}}).
\]
This stage uses the scene description as the source of narrative intent, the draft timeline as the base plan, and the speech results as trusted constraints. It may add, remove, or correct non-speech events when they are supported by the input scene. The refinement stage improves pacing, pauses, overlaps, and temporal coherence while preserving the relative order of semantically ordered events when required.

\subsection{Non-speech Generation and Post-production Planning}
After timestamp refinement, Audio-Oscar generates the non-speech audio elements specified in the refined timeline. Let
\[
\mathcal{E}_{\mathrm{ns}} = \mathcal{T} \setminus \hat{\mathcal{E}}_{\mathrm{sp}}
\]
denote the set of non-speech elements, including sound effects, music, and songs. For each element $z_i$, the system constructs an executable generation request.

Audio-Oscar employs specialized generation agents for different non-speech audio types. Given a planned non-speech element $z_i$, the corresponding agent selects an appropriate generation model and prepares model-specific generation parameters according to its audio type, textual description, target duration, and other information. The selected model then produces the audio
\[
\hat{y}_i = g_i(p_i, d_i, \rho_i),
\]
where $p_i$ is the generation prompt or description, $d_i$ is the target duration, and $\rho_i$ denotes auxiliary generation constraints.

Specifically, sound effects are handled by a text-to-audio generation agent, which converts the planned event description and duration requirement into a suitable prompt for a text-to-audio model. Music elements are processed by a music generation agent, which selects a suitable music model and prepares controls for style, mood and duration. Song elements are handled by a song generation agent, which prepares related controls such as lyrical content, vocal style, and accompaniment requirements before invoking the corresponding song generation model.

For sound effect generation, Audio-Oscar can further use an audio critic to evaluate generated candidates:
\[
\kappa_i=\mathcal{A}_{\mathrm{critic}}(z_i,\hat{y}_i),
\]
where $\kappa_i$ contains quality, semantic alignment, aesthetic scores, and suggestions. If a generated clip falls below a predefined threshold, the system uses critic feedback to revise the generation plan by rewriting the prompt, switching to a more suitable TTA model, or adjusting model parameters, and then retries generation for a bounded number of iterations.

Before final mixing, Audio-Oscar invokes an audio post-production planning agent to augment the refined timeline with post-production instructions. The agent analyzes the scene intent, temporal layout, and relationships among segments, and adds sparse post-production instructions for individual segments, such as gain adjustment, volume envelope, compression and reverberation. These instructions are consumed by the mixer to improve balance, transitions, spatial impression, and overall scene coherence.

\subsection{Mixing and Final Editing}
After speech and non-speech generation, Audio-Oscar combines all generated audio segments according to the timeline to produce a complete scene-level audio output.

Audio-Oscar performs a final editing stage after the initial mix. In this stage, the final mix review agent reviews the complete mixed audio in conjunction with the original scene description and the current timeline plan.

This stage targets scene-level issues that become apparent only after mixing, such as imbalanced loudness, abrupt transitions, masking between foreground and background elements, or insufficient spatial coherence. The final mix review agent proposes sparse per-segment remix patches to volume and edit metadata.
If valid final edits are produced, Audio-Oscar applies the patches to the audio plan and remixes the audio to obtain a refined final output. If no effective edit is needed, the initial mixed audio is kept as the final output.

\section{Experiment}

\begin{table*}[h]
    \centering
    \small
    
    \renewcommand{\arraystretch}{1}
    \adjustbox{max width=\textwidth}{
    \begin{tabular}{l ccc cccc}
    \toprule
    \multirow{2}{*}{Method} & \multicolumn{3}{c}{T2ABench} & \multicolumn{4}{c}{AudioTime} \\
    \cmidrule(lr){2-4} \cmidrule(lr){5-8}
    & Cnt-acc $\uparrow$ & Ord-acc $\uparrow$ & TS-acc $\uparrow$ & Ordering $\downarrow$ & Duration $\downarrow$ & Frequency $\downarrow$ & Timestamp $\uparrow$ \\
    \midrule
    AudioGen~\cite{kreuk2023audiogen}       & 5.40 & 6.00 & 18.40 & 0.91 & 3.73 & 1.58 & 0.54 \\
    AudioLDM~\cite{liu2023audioldm}       & 4.00 & 3.40 & 11.60 & 0.97 & 3.41 & 1.54 & 0.41 \\
    AudioLDM-2~\cite{liu2024audioldm}     & 7.40 & 1.20 & 13.40 & 0.96 & 3.40 & 1.64 & 0.54 \\
    Tango 2~\cite{majumder2024tango2}        & 4.60 & 10.20 & 18.80 & 0.86 & 3.70 & 1.52 & \underline{0.61} \\
    Make-An-Audio2~\cite{huang2023makeanaudio} & 4.00 & 19.80 & 18.80 & \underline{0.76} & 3.40 & 1.42 & 0.56 \\
    Stable Audio Open~\cite{evans2025stable} & 9.80 & 6.00 & \underline{21.80} & 0.98 & 3.07 & 1.46 & 0.53 \\
    MMAudio~\cite{cheng2025mmaudio}        & 4.80 & 2.40 & 21.40 & 0.98 & 3.33 & 1.54 & 0.50 \\
    AudioX~\cite{tian2026audiox}         & \underline{12.40} & \underline{23.60} & \textbf{28.20} & \textbf{0.34} & \textbf{1.30} & \textbf{0.74} & \textbf{0.81} \\
    \cellcolor{cyan!7}Audio-Oscar  & \cellcolor{cyan!7}\textbf{22.60} & \cellcolor{cyan!7}\textbf{69.80} & \cellcolor{cyan!7}20.20 & \cellcolor{cyan!7}0.79 & \cellcolor{cyan!7}\underline{1.44} & \cellcolor{cyan!7}\underline{1.08} & \cellcolor{cyan!7}0.54 \\

    \bottomrule
    \end{tabular}}
    \caption{
        Evaluation of instruction-following T2A ability on T2ABench and AudioTime. Except for our Audio-Oscar, all reported results are from AudioX~\cite{tian2026audiox}.
    }
    \label{tab:complex_tasks}
\end{table*}

\begin{table*}[h]
    \centering
    \small
    
    \renewcommand{\arraystretch}{1}
    \adjustbox{max width=\textwidth}{
    \begin{tabular}{l l l c cc ccc}
    \toprule
    \multirow{2}{*}{} &
    \multirow{2}{*}{Method} &
    \multirow{2}{*}{LLM} &
    \multirow{2}{*}{Thinking} &
    \multirow{2}{*}{Event (\%) $\uparrow$} &
    \multirow{2}{*}{Temporal (\%) $\uparrow$} &
    \multicolumn{3}{c}{LALM-based Score} \\
    \cmidrule(lr){7-9}
    & & & & & & Quality $\uparrow$ & Alignment $\uparrow$ & Aesthetic $\uparrow$ \\
    \midrule

    \multirow{7}{*}{Paired}
    & \cellcolor{gray!12}Reference Audio & \cellcolor{gray!12}-- & \cellcolor{gray!12}{--} & \cellcolor{gray!12}93.62 & \cellcolor{gray!12}95.26 & \cellcolor{gray!12}4.25 & \cellcolor{gray!12}4.54 & \cellcolor{gray!12}4.26 \\
    & Any2Speech & -- & -- & 80.38 & 84.54 & \underline{4.17} & 3.82 & 3.78 \\
    & WavJourney & DeepSeek-V4-Flash & \xmark & 87.26 & 92.09 & 3.84 & 3.90 & 3.57 \\
    & Audio-Oscar & DeepSeek-V4-Flash & \xmark & \textbf{92.14} & 93.36 &  4.13 & \underline{4.23} & \underline{3.96}  \\
    & Audio-Oscar & DeepSeek-V4-Flash & \cmark &  \underline{92.04} & \textbf{93.78} & 4.03 & 4.20 & 3.88 \\
    & Audio-Oscar & Qwen-122B-A10B & \xmark & 90.25 & 93.56 & \textbf{4.18} & \textbf{4.31} & \textbf{4.01} \\
    & Audio-Oscar & Qwen-397B-A17B  & \xmark & 91.31 & \underline{93.72} & 4.11 & 4.20 & 3.95 \\

    \midrule

    \multirow{6}{*}{Text}
    & Any2Speech & -- & -- & 51.65 & 41.16 & 3.53 & 2.63 & 2.74 \\
    & WavJourney & DeepSeek-V4-Flash & \xmark & 72.33 & 76.13  & 3.66 & 3.50 & 3.28 \\
    & Audio-Oscar & DeepSeek-V4-Flash & \xmark & \textbf{84.34} & \textbf{82.92} & \textbf{4.06} & \textbf{4.12} & \textbf{3.93} \\
    & Audio-Oscar & DeepSeek-V4-Flash & \cmark & 81.94 & 81.48 & 3.49 & 3.74 & 3.36 \\
    & Audio-Oscar & Qwen-122B-A10B & \xmark & 79.98 & 79.51 & \underline{4.04} & \textbf{4.12} & \textbf{3.93} \\
    & Audio-Oscar & Qwen-397B-A17B & \xmark & \underline{83.80} & \underline{81.79} & 3.95 & \underline{4.11} & \underline{3.86} \\

    \bottomrule
    \end{tabular}}
    \caption{
        Evaluation of event fidelity, temporal consistency, and LALM-based audio score on ASG-Bench. LLM denotes the large language model used as the backbone of each method. The best results are highlighted in bold, and the second-best results are underlined.
    }
    \label{tab:ASG-Bench}
\end{table*}

\subsection{ASG-Bench}

\begin{figure}[t]
  \centering
  \includegraphics[width=\linewidth]{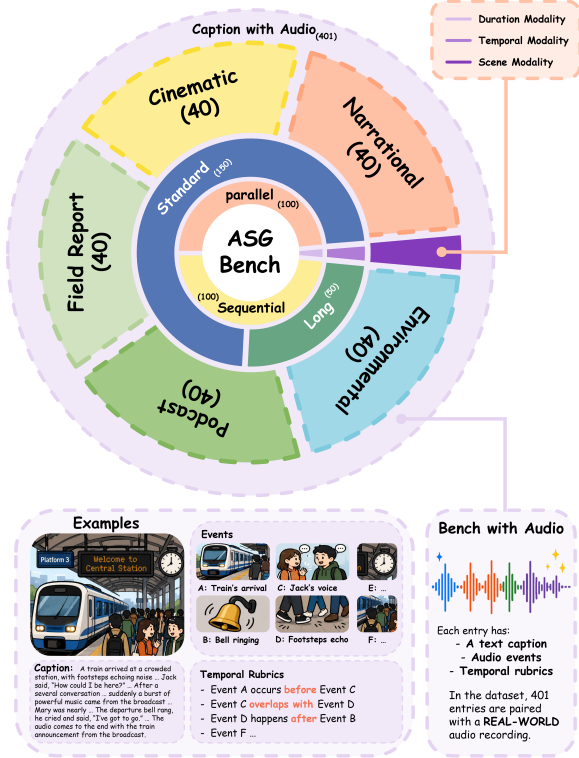}
  \caption{Overview of ASG-Bench, a benchmark for evaluating audio generation from complex audio scene descriptions. Each sample is annotated with expected audio events and temporal statements, supporting the assessment of whether generated audio faithfully captures both the scene content and temporal structure described in the prompt.}
  \label{fig:ASG-Bench}
\end{figure}

The effective evaluation of audio generation from complex audio scene descriptions remains a challenging problem, mainly due to the lack of benchmark datasets and the inherent difficulty of assessing generation quality. Based on this, we propose ASG-Bench, as illustrated in Figure~\ref{fig:ASG-Bench}. The dataset is composed of two parts: a reference-audio subset, where audio scene descriptions are paired with reference audio recordings, and a text-only subset, which contains audio scene descriptions without reference audio. Specifically, the reference-audio subset is constructed based on Omni-Cloze~\cite{omni-captioner}. In addition, each description is annotated with the expected audio events and temporal assertions, which are used for the preliminary evaluation of the quality of the generated audio. More details about this benchmark can be found in Appendix~\ref{apendix:ASG-Bench}.

\subsection{Experimental Setup}
\paragraph{Implementation Details}  In our experiments, we select a set of open-source models. Specifically, for the LLM-based agents, we use DeepSeek-V4-Flash~\cite{deepseekai2026deepseekv4}, Qwen3.5-122B-A10B, and Qwen3.5-397B-A17B~\cite{qwen3.5}. For the critic model, we employ a quantized version of Qwen3-Omni-30B-A3B-Instruct~\cite{xu2025qwen3}. For voice design, we use Qwen3-TTS-12Hz-1.7B-VoiceDesign~\cite{hu2026qwen3}. For text-to-speech synthesis, we use VoxCPM2~\cite{voxcpm2_2026,zhou2025voxcpm} and CosyVoice 3~\cite{du2025cosyvoice}; for text-to-audio generation, we adopt MMAudio~\cite{cheng2025mmaudio} and Stable Audio Open~\cite{evans2025stable}. For song generation, we use ACE-Step v1.5 with its 1.7B LM planner~\cite{gong2026ace}. In addition, we employ Qwen3-ForcedAligner-0.6B~\cite{shi2026qwen3} for precise timestamp alignment, and further use Demucs for vocal source separation in generated songs~\cite{rouard2022hybrid,defossez2021hybrid}. Except for the LLM-based agents accessed via official APIs, all the above models are deployed locally on RTX 4090 GPUs. In particular, Qwen3-TTS-VoiceDesign and VoxCPM2 are served using vLLM-Omni~\cite{yin2026vllmomni}. During inference, the temperature of all LLMs is set to 0.1. More details are provided in Appendix~\ref{ape:exp}

\paragraph{Benchmark and Metrics} 
For text-to-audio evaluation, we focus on its instruction-following capability. We therefore evaluate Audio-Oscar on T2A-bench~\cite{tian2026audiox} and AudioTime~\cite{xie2025audiotime}. 
These benchmarks focus on fine-grained controllability, including event counting, ordering, duration, frequency, and timestamp-related constraints. We follow the evaluation protocol used in AudioX and compare Audio-Oscar with the reported baseline results~\cite{tian2026audiox}.

To assess Audio-Oscar's ability to generate long-form audio that faithfully reflects complex audio scene descriptions, we evaluate it on ASG-Bench. We use Qwen3.5-Omni-Plus as the audio understanding evaluator~\cite{team2026qwen3}. Specifically, we first examine whether the audio events in each description are present in the generated audio, and whether the temporal statements are satisfied. We report the coverage rates of these events and temporal statements to measure event coverage and temporal consistency. In addition, we adopt a 1-5 rating scale and ask the evaluator to score each generated audio sample along three dimensions: audio quality for fidelity and listening clarity, scene alignment for consistency with the target description, and aesthetic appeal for overall naturalness and artistic quality. Detailed evaluation prompts and implementation details are provided in Appendix~\ref{apendix:ASG-Bench}.

\subsection{Main Results}

\paragraph{Instruction-following text-to-audio generation}
Table~\ref{tab:complex_tasks} reports the results on T2A-Bench and AudioTime. To ensure consistency across experiments, we use Stable Audio Open as the only TTA generation model. On T2A-Bench, Audio-Oscar achieves the best Cnt-acc and Ord-acc, showing the benefit of explicit scene planning for compositional event organization. It also obtains near-perfect category accuracy, since event categories are explicitly parsed and routed during planning; we therefore focus the main table on counting, ordering, and timestamp metrics. However, its TS-acc remains below AudioX, indicating that precise timestamp control is still limited by the underlying T2A generator. On AudioTime, Audio-Oscar obtains competitive results on duration and frequency metrics. This suggests that planning-based decomposition helps with temporal attributes such as duration and repetition. Nevertheless, it still lags behind AudioX on ordering and timestamp metrics, showing that fine-grained temporal grounding remains a limitation. Moreover, compared with the original results of Stable Audio Open on both benchmarks, our method achieves clear improvements, further showing the effectiveness of our system.

\paragraph{Complex Audio Scene Generation}
Table~\ref{tab:ASG-Bench} reports the results on ASG-Bench. We compare Audio-Oscar with WavJourney~\cite{liu2025wavjourney} and Any2Speech~\cite{song2026borderless}. For WavJourney, we use DeepSeek-V4-Flash in non-reasoning mode. More experimental details can be found in Appendix~\ref{ape:exp}. We further present the performance of Audio-Oscar when different LLMs are used as the agent. As shown by the experimental results, Audio-Oscar achieves great performance on ASG-Bench when using different LLMs. For the subset with reference audio, Audio-Oscar achieves audio-event and temporal-statement scores that are close to those of the reference audio across multiple LLM settings, demonstrating its ability to faithfully capture target scene events and preserve temporal structure. Meanwhile, Audio-Oscar also achieves higher LALM-based scores than the baseline systems, approaching the quality of reference audio. On the more challenging text-only subset, Audio-Oscar also achieves strong results across different LLM backbones and clearly outperforms the compared methods.
We also observe that enabling the thinking mode for DeepSeek does not improve performance. This may suggest that disabling the thinking mode can reduce inference cost while maintaining competitive performance in our pipeline.

\paragraph{Human Evaluation}
We conduct a human evaluation to assess the quality of the audio generated by our method. Specifically, we perform a subjective listening study using mean opinion scores (MOS). Human listeners evaluate each audio sample along three dimensions: Quality, Alignment, and Aesthetic. They are asked to provide an overall score for each generated audio sample. We randomly sample 10 examples from each subset of ASG-Bench, resulting in a total of 20 audio samples for human evaluation. We collect scores from 10 participants, with each participant asked to rate the samples on a 1–5 scale. More details are provided in Appendix~\ref{app:human-eval-prompt}. As shown in Figure~\ref{fig:subject}, our method also achieves strong performance in the subjective evaluation. Compared with other methods, our method obtains clearly higher subjective scores, demonstrating its ability to generate audio scenes that better align with human subjective preferences. 
\begin{figure}[t]
  \centering
  \includegraphics[width=\linewidth]{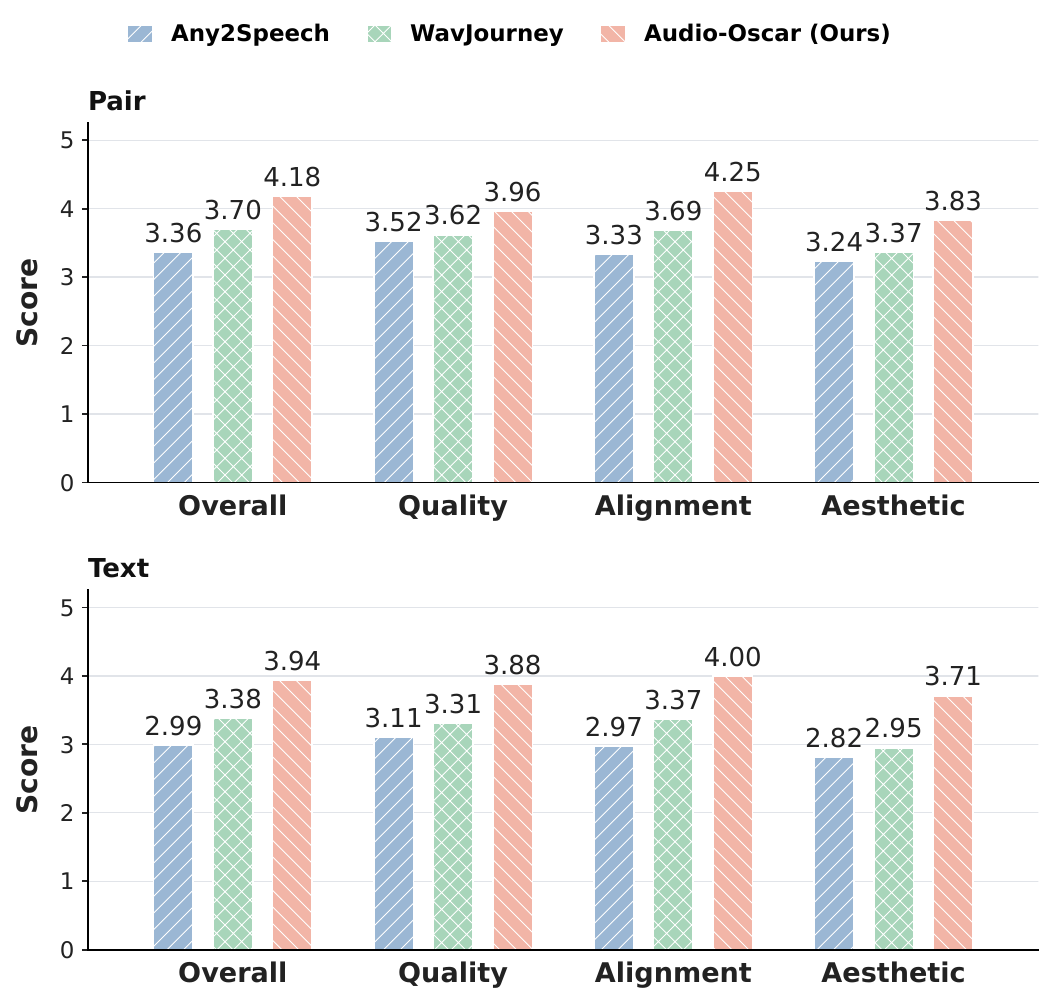}
  \caption{Human Subjective Evaluation Scores for Sampled Examples from ASG-Bench. WavJourney and our method are evaluated using DeepSeek-V4-Flash with reasoning mode disabled. }
  \label{fig:subject}
\end{figure}

\begin{figure}[t]
  \centering
  \includegraphics[width=\linewidth]{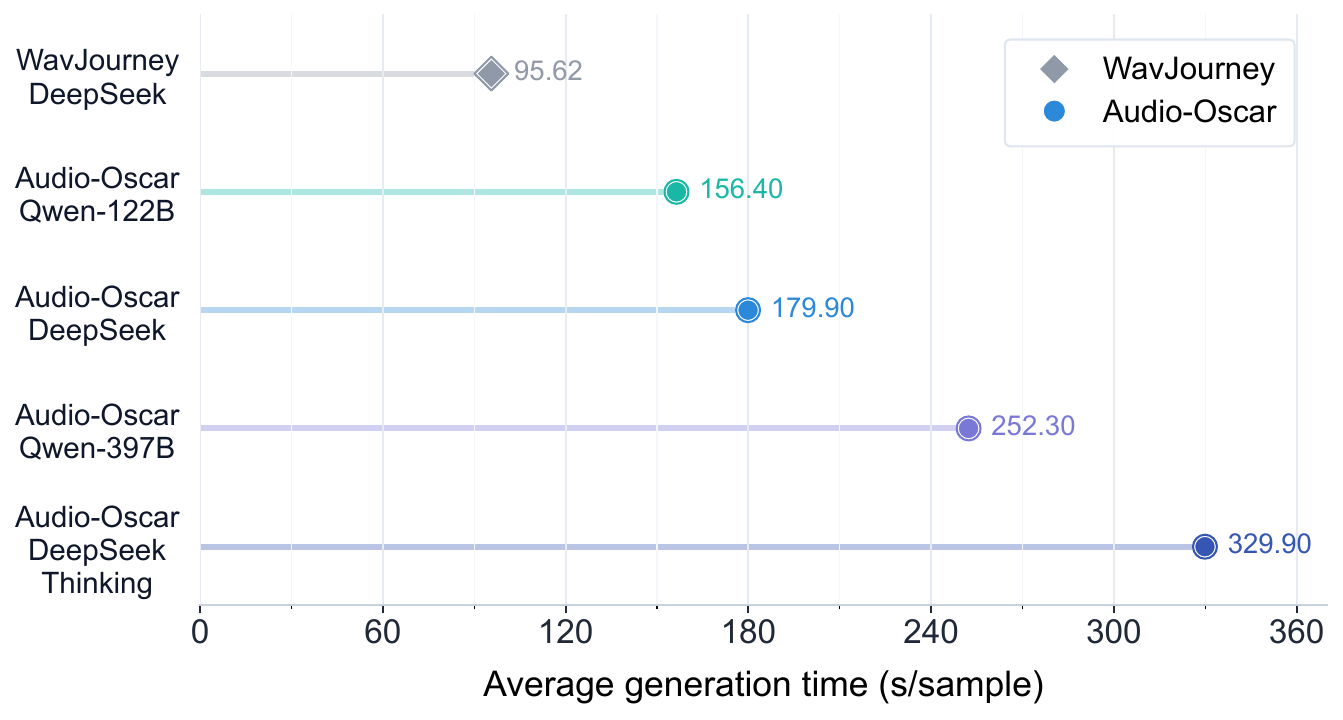}
  \caption{Average generation time of WavJourney and Audio-Oscar on randomly sampled ASG-Bench examples. DeepSeek refers to DeepSeek-V4-Flash. Unless otherwise specified, all LLMs are used in non-thinking mode.}
  \label{fig:time}
\end{figure}

\paragraph{Runtime Analysis} We further evaluate the runtime of Audio-Oscar by randomly sampling 10 examples from ASG-Bench and measuring the average generation time under different LLMs. As shown in Figure~\ref{fig:time}, Audio-Oscar requires 156.40--329.90 seconds per sample depending on different LLMs, whereas WavJourney requires 95.62 seconds per sample. The additional runtime is expected as Audio-Oscar performs role modeling and voice design, timeline refinement, critic-guided repair and post-production planning. Since runtime depends on scene complexity, the generated timeline structure, API latency, and generation models, these results are intended only as a reference.

\section{Conclusion}
In this work, we introduce Audio-Oscar, a multi-agent system designed for complex audio scene generation. Starting from an audio scene description, Audio-Oscar decomposes generation into a structured production workflow, including role modeling and voice design, event and timeline planning, timeline scheduling, model orchestration, and audio post-production. By integrating structured planning, model orchestration and audio post-production into a unified pipeline, Audio-Oscar is able to handle audio scenes involving dialogue, environmental sounds, music, songs, and complex temporal relationships. We also introduce ASG-Bench. The benchmark is designed around complex audio scene descriptions and aims to assess whether a model or system can accurately understand and generate final audio that is consistent with the input description. Our experimental results show that Audio-Oscar can effectively generate complex audio scenes.

\label{sec:bibtex}

\section*{Limitations}
Audio-Oscar uses a Voice-Design TTS model in its pipeline to create and assign customized voices for different characters, improving consistency and controllability in multi-character speech generation. However, compared with real recorded human voices, these generated voices may still be limited in authenticity, naturalness, and subtle emotional expressiveness. 

In addition, when handling especially long audio requests, the planning stage of Audio-Oscar may sometimes arrange the content as a single long generation segment. Such a segment can exceed the stable capability range of the underlying generation model, which may affect temporal alignment and overall audio quality. For certain transient or extreme audio events, such as door slams, impacts, or explosions, the planner may also assign an overly short duration. If the duration is too short, the generation model may struggle to capture the full attack, body, and decay of the sound, resulting in audio that is less clear or realistic. 

Finally, for the evaluation phase, our benchmark relies on large models for rubric-based scoring. Thus, the reliability of our evaluation is currently bounded by the capabilities of the evaluator models, though its effectiveness will steadily scale and improve alongside future advancements in audio foundation models.

\section*{Ethics Statement}
This work aims to generate long-form audio that better aligns with users’ scenario descriptions by leveraging multi-agent collaboration to autonomously plan the audio generation pipeline, including timeline arrangement, the selection and invocation of different open-source models, and necessary audio post-production. We currently use open-source models and encourage users to carefully consider potential copyright issues when using the system. Given the risk of misuse, users are expected to ensure that both their input content and the generated outputs comply with applicable norms and ethical standards.


\bibliography{custom}

\appendix

\section{Audio-Oscar}
\label{apd:audiooscar}
In Audio-Oscar's pipeline, following prior related work~\cite{liu2025wavjourney,rong2025audiogenie}, we consistently employ structured JSON as the output format and use it to pass information across different stages of the system. 

\subsection{Agents in Audio-Oscar}
\paragraph{Scene Intake Agent} The Scene Intake Agent is an optional module that converts user input into a canonical scene description for downstream audio planning. The agent supports two modes. In script mode, it conservatively normalizes an existing audio description by improving punctuation, paragraphing, sentence flow, and clarity, while strictly preserving the original language as well as all explicitly stated sound events, dialog and constraints. No deletion, modification, or addition is allowed. In idea mode, the agent expands a brief user idea into a fuller audio scene description by making only reasonable and scene-consistent elaborations, thereby providing downstream agents with a generation-ready description. In our evaluation, we use only the script mode, while the idea mode is included solely as an extended functionality.

\paragraph{Speaker Role Profiling Agent} The Speaker Role Profiling Agent extracts speaking roles from the scene description and constructs role profiles. For each detected role, the agent outputs a role identifier, a role description, a stable voice description, and a short sentence for zero-shot voice reference. For voice descriptions, we require the model to focus primarily on stable acoustic attributes such as gender, age, pitch, speaking rate, articulation clarity, and accent, while avoiding scene-specific emotions or utterance-level performance variations. These role profiles provide the basis for speaker identity modeling and voice design in subsequent TTS generation.

\paragraph{Speech Event Planning Agent} The Speech Event Planning Agent generates a complete speech plan required for subsequent TTS generation based on the scene description and the extracted speaker profiles. It is responsible for identifying all speech events in the input and assigning each utterance to its corresponding speaking role. The agent is constrained to preserve the original dialogue and is not allowed to create new dialogue or rewrite existing utterances. The final output is a structured list of speech segments, where each segment contains the speaker name and the exact text to be synthesized.

\paragraph{Audio Timeline Planning Agents} First, the Audio Timeline Drafting Agent determines which non-speech events are needed in the scene and assigns each segment an audio type, acoustic description, start time, duration, and initial volume. It also arranges non-speech sounds around the generated speech segments. Speech segments are treated as fixed events, their speaker names, text, and durations are preserved, and the agent is only responsible for scheduling their start time and end time. In addition, the agent marks speech segments that require forced alignment when necessary, so that finer-grained word-level timing information can be obtained in cases involving nearby detailed sound effects, interruptions, or overlapping events. The Timestamp Refinement Agent then repairs and refines the draft timeline. It takes the scene description, the draft timeline, trusted speech durations, and available forced-alignment information as input, and improves pacing, pauses, overlaps, backgrounds, and temporal coherence. The agent may only add, remove, or correct non-speech events supported by the scene description, while ensuring that every speech segment is fully preserved and appears exactly once. The final output is a complete audio timeline.

\paragraph{Non-speech Generation Planning Agents} The agents convert the audio timeline into executable generation plans for different types of non-speech audio. Follow \citet{rong2025audiogenie}, Audio-Oscar assigns expert agents to sound effects, music, and songs, respectively, with each agent processing only the segments of the corresponding type in the timeline. The TTA Generation Planning Agent is responsible for sound effects. For each segment, it selects an appropriate TTA model and generation parameters, and reformulates the segment description into a suitable generation prompt to the model. When critic feedback is available, this agent can also repair generation results with low scores by modifying the prompt, switching models, or adjusting parameters, but it is not allowed to alter the timeline or the semantic meaning of the events. The Music Generation Planning Agent handles non-lyrical music segments, such as background music. It selects suitable models and parameters for each music segment and converts the segment into a complete, generatable music description. Meanwhile, it preserves all timeline-related information. The Song Generation Planning Agent is responsible for song segments that require singing or lyrical content. For each song segment, it selects an appropriate song generation model and parameters, and produces the style tags, lyrics, or singing descriptions required by the song generation model. Like the other generation planning agents, it strictly preserves the timeline and only completes the model inputs required for song generation.

\paragraph{Audio Critic and TTA Repair Agent} The Audio Critic and the TTA Repair Agent provides feedback-driven quality control and iterative refinement for generated sound effects. Given a planned audio event and its generated audio clip, the Audio Critic listens to the clip and scores it from 0 to 1 along three dimensions: quality, which measures audio fidelity, clarity, naturalness; alignment, which measures whether the generated audio matches the target audio description; and aesthetics, which measures the usability and pleasantness of the clip. If any score falls below a predefined threshold, the critic returns revision suggestions. The TTA Repair Planner then uses this feedback to revise low-scoring generation plans by rewriting the generation prompt, switching to a more suitable TTA model, or adjusting model parameters. This enables iterative improvement of low-quality or semantically misaligned sound effects.

\paragraph{Audio Post-production Planning Agent} Audio-Oscar invokes an Audio Post-production Planning Agent to supplement the refined audio timeline with segment-level editing instructions. This agent analyzes the scene, temporal layout, and relationships among segments, and attaches sparse post-production metadata to individual segments. Currently, The supported operations include gain adjustment, fade-in and fade-out, crossfade, volume envelope, ducking, panning, equalization, compression, and reverberation. This agent only specifies how each generated segment should be shaped in the final mix.

\paragraph{Final Mix Review Agent} After the initial mix is generated, Audio-Oscar can invoke a Final Mix Audit Agent to perform an additional review pass. This agent is based on an audio-capable omni model. It listens to the complete mixed audio and compares it with both the planned scene intent and the actual mixer manifest. The agent receives segment-level context that includes the original plan and the mixer-applied results, and identifies only actionable mix-level issues, such as abrupt boundaries, inappropriate dryness or reverberation, and poor loudness balance. It outputs sparse per-segment remix patches only for existing segments. The allowed updates are limited to volume adjustments and post-production edit fields. The mixer then applies these patches to produce an optimized final audio output, improving the balance, transitions, and overall coherence of the final mix.

\label{sec:appendix}

\section{Experiments}
\label{ape:exp}
\subsection{Implementations Configuration}
When evaluating T2ABench and AudioTime, we skip the Scene Intake Agent to ensure consistency across inputs. In contrast, for ASG-Bench, we retain the use of the Scene Intake Agent. 
During evaluation on ASG-Bench, we use the following models: for voice design, we select Qwen3-TTS-12Hz-1.7B-VoiceDesign~\cite{hu2026qwen3}. For sound effects generation, we use MMAudio with the \texttt{large\_44k\_v2} variant~\cite{cheng2025mmaudio} and Stable Audio Open with the \texttt{stabilityai/stable-audio-open-1.0} checkpoint~\cite{evans2025stable}, both using their official default inference parameters. For music generation, considering that the required audio is primarily background or instrumental music, we use the same Stable Audio Open checkpoint as before~\cite{evans2025stable}. For song generation, we use ACE-Step v1.5 with its 1.7B LM model for planning and base DiT model for execution~\cite{gong2026ace}. In addition, we continue to use Qwen3-ForcedAligner-0.6B~\cite{shi2026qwen3}, Demucs~\cite{rouard2022hybrid,defossez2021hybrid} and Qwen3-Omni-30B-A3B-Instruct-AWQ-4bit~\cite{xu2025qwen3}. For the use of Demucs, we introduce an additional parameter, \texttt{pure\_vocal}, into the model-selection parameters of ACE-Step for Song Generation Planning Agent. When this parameter is set to \texttt{true} by the agent, the generated song is automatically routed to Demucs for post-processing, where the vocal stem is separated from the accompaniment to obtain a vocal-only output. For critic-guided TTA repair, we set the repair threshold to 0.7 and allow at most three repair retries. 

When evaluating T2ABench and AudioTime, we use only Stable Audio Open~\cite{evans2025stable} for sound effects generation to ensure comparability. Meanwhile, we keep all other pipelines and models unchanged, so as to evaluate the performance of our system under realistic deployment settings.
\subsection{Details of Any2Speech Evaluation}
\label{sec:any2speech-eval}
To the best of our knowledge, Any2Speech~\cite{song2026borderless} did not provide a public API at the time our experiments were conducted. We therefore evaluated it through its publicly available web interface. Any2Speech provides a scene-setting option in its interface. To make the comparison fair and to prevent the system from directly reading environmental descriptions as audiobook-style narration, we prepend a short task instruction to each benchmark prompt: "\textit{Generate vivid scenes, not an audiobook.}" The original scene description from ASG-Bench is then appended after this instruction. We collect the generated audio outputs for all 601 examples in ASG-Bench.

This protocol allows us to compare the final audio quality and the instruction-following ability of Any2Speech with other systems. However, it does not provide a reliable measurement of generation latency. Since Any2Speech is accessed through a web interface rather than a controllable local or programmatic API, the observed runtime may include request scheduling, server-side availability, network transmission, and interface-related overhead. These factors are not directly comparable with the controlled execution time of Audio-Oscar and WavJourney. Therefore, we exclude Any2Speech from the time consumption comparison.

\subsection{Details of WavJourney Evaluation}
\label{sec:wavjourney-eval}
While WavJourney~\cite{liu2025wavjourney} provides a comprehensive open-source framework for evaluation, we introduced several modifications to the original pipeline to ensure fairness and architectural consistency during benchmarking.

First, we upgraded the foundational models to align with modern baselines. For TTS, the original HuBERT-Bark~\cite{hsu2021hubert, Bark} pipeline was replaced with CosyVoice3 to enable direct reference audio input and superior synthesis quality. For TTA and TTM, we consolidated the separate AudioGen~\cite{kreuk2023audiogen} and MusicGen~\cite{copet2023simple} components into Stable Audio Open. Furthermore, the central LLM for planning was switched from GPT-4~\cite{achiam2023gpt} to DeepSeek-V4-Flash in non-reasoning mode.

Second, we adapted the voice timbre matching logic. Since WavJourney requires reference audio for speech synthesis, we utilized the audio generated during the voice design phase of our Audio-Oscar framework as the reference timbre library. Because both frameworks employ LLMs for dynamic persona planning, discrepancies often arise in character counts and naming conventions. To resolve this, we implemented an LLM-based semantic alignment mechanism that performs matchings between WavJourney’s planned target characters and our timbre library, guaranteeing consistent voice assignment for every role. To ensure a fair comparison, the timbre library for each setting is also generated by our framework using DeepSeek-V4-Flash in non-reasoning mode, matching the planning LLM configuration used in the adapted WavJourney pipeline.

Additionally, we resolved several operational anomalies in the original pipeline, such as language mismatches and invalid planning parameters. These issues were mitigated through prompt optimization and runtime dynamic validation. Ultimately, we successfully evaluated the modified WavJourney framework on 601 samples from the ASG-Bench dataset, collecting both the generated audio and their corresponding time consumption metrics for empirical comparison.

\subsection{Instruction-Following Text-to-Audio Benchmarks}
\label{sec:audiotime_t2abench}
To further evaluate the instruction-following ability of Audio-Oscar for controllable text-to-audio generation, we conduct additional experiments on T2A-bench~\cite{tian2026audiox} and AudioTime~\cite{xie2025audiotime}. T2A-bench is designed to evaluate fine-grained text-to-audio controllability with natural-language prompts, covering category generation, event counting, event ordering, and timestamp control. AudioTime focuses on temporal audio-text alignment and evaluates whether the generated audio follows temporal instructions involving ordering, duration, frequency, and timestamps.

For these benchmarks, we evaluate the non-speech audio generation pathway of Audio-Oscar. Given each benchmark prompt, the system treats it as a sound-event generation request and generates the corresponding audio scene. We adhere to the official evaluation protocols of the two benchmarks. For T2A-bench, we report count accuracy, ordering accuracy, and timestamp accuracy. We omit the category subset of T2A-bench from the main table. This subset mainly evaluates whether the generated audio contains the sound category specified in the prompt. Since Audio-Oscar explicitly extracts the target category during the planning stage before routing the request to the audio generation module, evaluating this subset would render the task trivial, failing to provide a rigorous and meaningful assessment of the model's inherent instruction-following behavior. For AudioTime, we report the STEAM metrics for ordering, duration, frequency, and timestamp control. All generated audio clips are saved with the file names required by the official evaluation scripts, and no reference audio from the test sets is used during generation.

\subsection{Details of Human Subjective Evaluation}
\begin{figure}[t]
    \centering
    \includegraphics[width=\columnwidth]{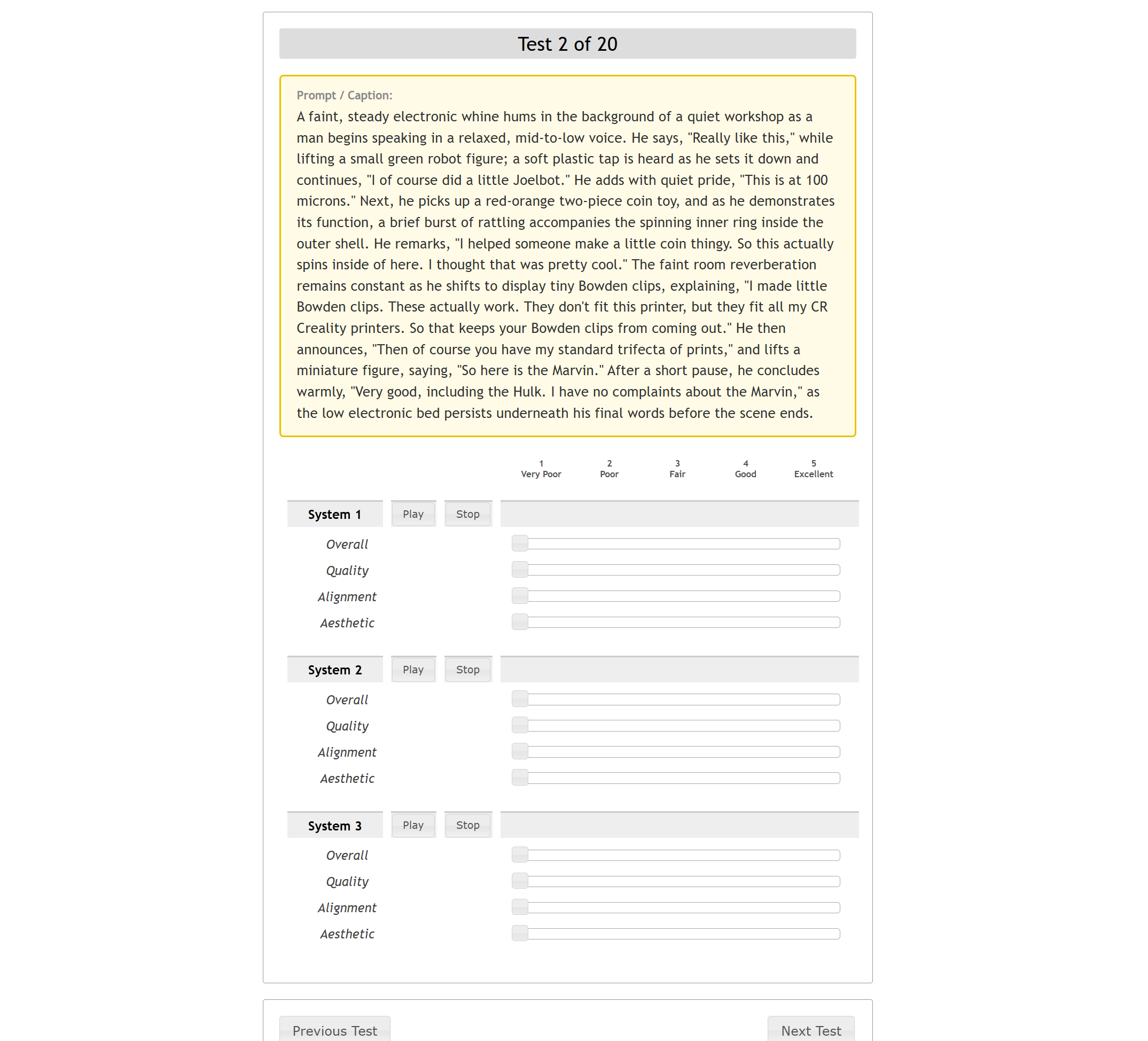}
    \caption{Illustration of the human evaluation platform. Participants are shown a text prompt and three generated audio samples, and are asked to rate each sample along four dimensions.}
    \label{fig:eval-platform}
\end{figure}

We conducted a subjective listening study to assess the audio generated by our method. To conduct the subjective evaluation, we sampled 10 examples from each subset of ASG-Bench, resulting in a total of 20 audio samples for human evaluation. For the text-only subset, we leveraged its comprehensive categorization to construct a carefully balanced sample set, including two examples per scene modality, five examples per temporal modality, five examples per language, and a mixture of four long and six short audio clips. For the subset with reference audio, we randomly sampled 10 examples for evaluation. 
Human listeners evaluated each audio sample along three dimensions: Quality, Alignment, and Aesthetic, and also provided an overall score. The instructions shown to participants are presented below.  An illustration of the evaluation platform is shown in Figure~\ref{fig:eval-platform}.
\label{app:human-eval-prompt}

\definecolor{PromptBack}{HTML}{F8FAFF}
\definecolor{PromptFrame}{HTML}{AAB7D8}
\definecolor{PromptTitleBack}{HTML}{EEF3FF}
\definecolor{PromptTitleText}{HTML}{1F2A44}

\newtcblisting{insturct}[2][]{
  enhanced,
  breakable,
  listing only,
  colback=PromptBack,
  colframe=PromptFrame,
  colbacktitle=PromptTitleBack,
  coltitle=PromptTitleText,
  title={#2},
  fonttitle=\bfseries,
  boxrule=0.45pt,
  arc=1.2mm,
  left=6pt,
  right=6pt,
  top=6pt,
  bottom=6pt,
  listing options={
    basicstyle=\ttfamily\small,
    breaklines=true,
    breakatwhitespace=true,
    breakautoindent=false,
    breakindent=0pt,
    columns=fullflexible,
    keepspaces=true
  },
  #1
}

\begin{insturct}{Instructions for Participants}
Thank you for participating in this audio evaluation. Please read the instructions carefully before starting.

Instructions:
- Use high-quality headphones and a good soundcard.
- Each page shows a text prompt describing a scene, followed by three audio samples generated by different systems.
- Listen to all three samples carefully. You may replay them as many times as needed.
- Rate each sample on the following four dimensions using a 1--5 scale:

  Overall: Your overall impression of the audio quality.
  Quality: Audio fidelity, clarity, noise level, distortion/artifacts, intelligibility, and listening comfort.
  Alignment: Semantic, temporal, spatial, and scene-coherence consistency between the generated audio and the caption/input scene.
  Aesthetic: Overall artistic appeal, naturalness of the sound design, emotional impact, and pleasantness.

Scoring scale:
1 = Very Poor
2 = Poor
3 = Fair
4 = Good
5 = Excellent
\end{insturct}

\section{ASG-Bench}
\label{apendix:ASG-Bench}
\subsection{Overview} 
Evaluating audio generation models against complex, naturalistic audio scene descriptions presents two challenges: the scarcity of dedicated benchmark datasets and the inherent difficulty of assessing generation quality in the audio domain. 
Existing benchmarks either focus on isolated sound events, single-track speech synthesis, or music generation, leaving a critical gap in the evaluation of models capable of producing multi-track, temporally rich audio scenes that interweave speech, sound effects, music, and environmental ambience.

To address this gap, we introduce \textbf{ASG-Bench}, a comprehensive benchmark dataset comprising 601 audio scene descriptions. The dataset is organized into two complementary subsets: \textbf{Captions with Audio Files}, containing 401 entries each paired with a real-world audio recording, and \textbf{Text-only Caption}, containing 200 text-only descriptions, evenly distributed between Chinese and English (100 each). Each entry is meticulously annotated with expected audio events and temporal assertions, enabling preliminary yet structured evaluation of generated audio quality.

\subsection{Dataset Taxonomy}\label{sec:taxonomy}
\subsubsection{Audio Scene Caption (200 Items)}\label{sec:asc_taxonomy}
The Audio Scene Caption subset is constructed along four independent dimensions: scene modality, temporal modality, language, and duration. The design ensures linear independence across all dimensional combinations, yielding a balanced and comprehensive evaluation corpus.

\paragraph{Scene Modality} We categorize audio scenes into five modality classes, with 40 entries per class. \textbf{Cinematic} scenes resemble film or action drama excerpts, typically involving rich sonic layers such as dialogue, ambient effects, and background score. \textbf{Narrational} scenarios are characteristic of audiobooks or narrated documentaries, where a primary speaker describes or tells a story. \textbf{Environmental} entries capture everyday or natural environments (e.g., street markets, forests, offices) where ambient sound dominates. \textbf{Podcast/Broadcast} refers to formal or semi-formal spoken content with moderate production, often involving multiple speakers or call-in formats. \textbf{Field Reporting} designates on-the-spot recordings of real-world events, capturing authentic and unscripted acoustic environments.

\paragraph{Temporal Modality} We identify two temporal patterns, \textbf{Parallel} and \textbf{Sequential}, with 100 entries per class. In parallel class, multiple audio events occur concurrently and interweave, testing the model's ability to orchestrate foreground-background mixing. In sequential class, audio events occur in a clearly ordered manner with distinct transitions, testing timestamp planning and event alignment.

\paragraph{Language} The corpus is bilingual, with 100 entries in \textbf{Chinese} and 100 in \textbf{English}. The two language subsets are balanced across all other dimensional categories.

\paragraph{Duration} We define two duration regimes, \textbf{Standard} and \textbf{Long}. For caption-based audio duration estimation, standard audio clips ranging from 10 to 20 seconds while long audio clips ranging from 45 to 60 seconds. This is specifically reflected in the caption as the size of the text length and the number of speech events. Note that Long-duration entries cannot be Parallel, as the concurrent interweaving of multiple audio streams inherently constrains temporal duration---maintaining coherent foreground-background interleaving over 45--60 seconds is practically infeasible without temporal segmentation, which would violate the Parallel definition.

\subsubsection{Audio Scene Caption with Reference Audios}\label{sec:ascf_taxonomy}

The subset with reference audios of ASG-Bench is constructed based on Omni-Cloze~\cite{omni-captioner}. Specifically, we first extract audios from the corresponding Omni-Cloze videos to obtain real audio files, and then filter the resulting clips by retaining only those with durations between 20 and 60 seconds. We then replace the blanks in the original passage with their correct answers to get reference video descriptions. To prioritize samples with richer audio information, we count the number of cloze questions involving the audio modality in each sample, sort the samples in descending order accordingly. Then, we use WhisperX to automatically transcribe the extracted audio, providing speech-content references for subsequent audio scene caption generation~\cite{bain2023whisperx}. Finally, we employ Qwen3.5-Omni-Plus to generate the audio descriptions~\cite{team2026qwen3}. The model takes the reference video description, the extracted audio, and the transcribed speech as input, and produces the final audio scene description.

\begin{figure}[t]
  \centering
  \includegraphics[width=\linewidth]{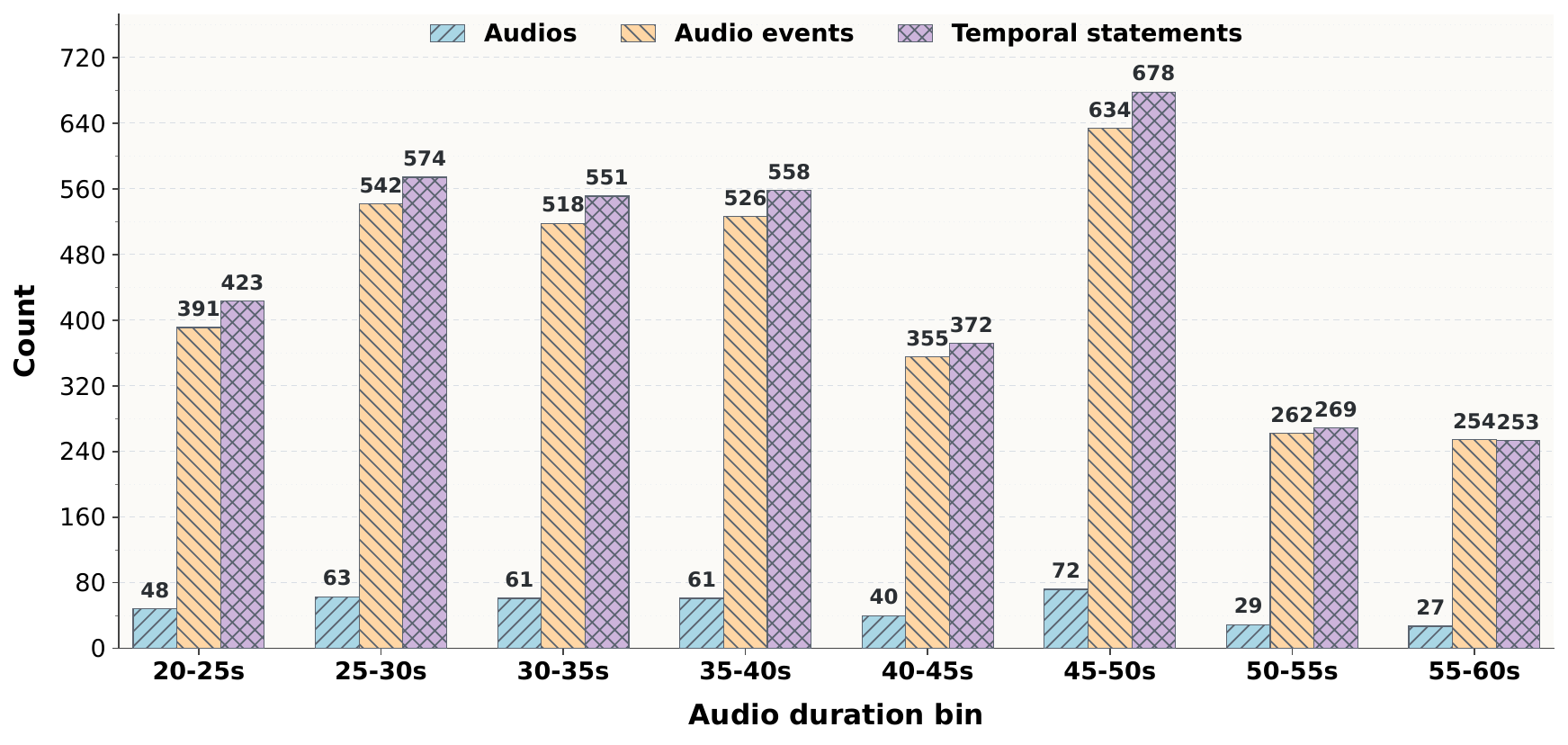}
  \caption{Distribution of Audios, Audio Events and Temporal Statements Across Duration Intervals}
  \label{fig:duration}
\end{figure}

After obtaining the audio scene descriptions, we further use an LLM to derive the audio events that should be present in the audio, as well as temporal statements that the audio should satisfy. Specifically, we use DeepSeek-V4-Pro for this annotation~\cite{deepseekai2026deepseekv4}. We then conduct a round of human annotation, where annotators subjectively assess the consistency between each audio scene description and the corresponding audio, assigning a score from 1 to 5. The annotators also verify the correctness of each extracted audio event and temporal statement. Finally, we filter out samples with consistency scores below 4 and remove audio events and temporal statements judged to be incorrect. The distribution of audio clips, temporal assertions, and audio events across different duration intervals is shown in Figure~\ref{fig:duration}.

\subsection{Content Structure}\label{sec:content_structure}
Each entry in ASG-Bench contains three core components: a \textbf{text caption}, a set of \textbf{audio events}, and a set of \textbf{temporal statements}. The Audio Scene Caption with Files subset additionally includes a reference audio file. An example is shown in Table~\ref{tab:ASG-Bench_example}.

\paragraph{Text Caption} The caption is a coherent natural-language description of an audio scene. A key annotation principle is that temporal information must be embedded naturally and unobtrusively: over-explicit phrasing (e.g., \textit{"the wind starts exactly when Jack finishes and ends when singing begins"}) encourages rigid pattern matching but ignores natural expression, while complete omission (e.g., \textit{"Jack is speaking, the wind is blowing, someone is singing"}) leaves the model unable to infer event relationships. A well-formed caption naturally integrates temporal cues: \textit{"Jack finished speaking, then a gust of wind swept through, followed by the sound of singing."}

\paragraph{Audio Events} Each caption yields a set of atomic audio events (e.g., spoken lines, wind, music). Events serve as the primary evaluation metric, measuring whether the generated audio reproduces the expected sound content.

\paragraph{Temporal Assertions} Relationships between events are captured via three assertion types: A after B, A before B, and A overlaps with B. These constitute a secondary evaluation metric, assessing whether the generated audio preserves the intended temporal structure.

\begin{table*}[t]
\centering
\small
\begin{tabular}{p{0.18\textwidth} p{0.76\textwidth}}
\toprule
\textbf{Component} & \textbf{Example} \\
\midrule
Text Caption &
A steady, rhythmic whoosh of ocean waves crashing onto a shore establishes a vast open space from the outset. The narrator's voice, resonant and old, begins: "The lighthouse keeper had not seen a ship in weeks." The cry of a distant seagull pierces the air, its call echoing. He continues, "The storm was coming; he could feel it in his bones." A heavy gust of wind rattles the windows of the lighthouse, a creaking sound of old wood straining. He says, "He climbed the spiral stairs, his boots heavy on the iron." Each step is a metallic clang, echoing in the stone tower. As he reaches the top, the wind howls more fiercely, blending with the waves. He adds, "He lit the lamp, a single beam cutting through the dark." A soft click of a switch, followed by a low electrical hum, fills the space. The waves crash, the wind groans, and the hum persists. He says, "It was all he could do." The sounds of nature and the lighthouse continue, a symphony of solitude. \\

\midrule
Audio Events &
\begin{tabular}[t]{@{}l@{}}
Steady, rhythmic whoosh of ocean waves crashing onto a shore.\\
      Resonant and old narrator says "The lighthouse keeper had not seen a ship in weeks."\\
      Cry of a distant seagull, echoing.\\
      Resonant and old narrator says "The storm was coming; he could feel it in his bones."\\
      A heavy gust of wind rattling windows, accompanied by the creaking of old wood straining.\\
      Resonant and old narrator says "He climbed the spiral stairs, his boots heavy on the iron."\\
      A series of metallic clangs as boots step on iron stairs, echoing in a stone tower.\\
      Wind howling fiercely, blending with the waves.\\
      Resonant and old narrator says "He lit the lamp, a single beam cutting through the dark."\\
      A soft click of a switch, followed by a low electrical hum filling the space.\\
\end{tabular} \\

\midrule
Temporal Statements &
\begin{tabular}[t]{@{}p{0.98\linewidth}@{}}
The rhythmic whoosh of ocean waves crashing overlaps with the narrator's speech ``The lighthouse keeper had not seen a ship in weeks.'' \\
The cry of a distant seagull is heard after the narrator says ``The lighthouse keeper had not seen a ship in weeks.'' \\
The sound of a heavy gust of wind rattling windows and creaking old wood occurs after the narrator says ``The storm was coming; he could feel it in his bones.'' \\
The narrator says ``He climbed the spiral stairs, his boots heavy on the iron.'' after the wind rattling windows and creaking sound. \\
The metallic clanging of footsteps on iron stairs begins after the narrator says ``He climbed the spiral stairs, his boots heavy on the iron.'' \\
The wind howls more fiercely after the metallic clanging of footsteps. \\
The narrator says ``He lit the lamp, a single beam cutting through the dark.'' after the wind howls more fiercely. \\
The soft click of a switch is heard after the narrator says ``He lit the lamp, a single beam cutting through the dark.'' \\
The low electrical hum begins after the soft click of a switch. \\
The low electrical hum overlaps with the narrator saying ``It was all he could do.''
\end{tabular}\\
\bottomrule
\end{tabular}
\caption{An example entry in ASG-Bench. Each entry consists of a natural-language text caption, a set of audio events, and temporal statements describing relationships among events.}
\label{tab:ASG-Bench_example}
\end{table*}

\subsection{Evaluation Methodology}\label{sec:evaluation}
ASG-Bench uses Qwen3.5-Omni-Plus as the audio understanding evaluator~\cite{team2026qwen3}. 
Given a generated audio clip, the evaluator is provided with the target caption and the corresponding evaluation annotations, and is instructed to judge only what is actually audible in the generated audio. All evaluations in this work are conducted through the official API, with the temperature set to 0.1. Each sample is evaluated once. Since the evaluation is performed via the official API, a small number of samples may fail during evaluation because the model output is flagged as containing harmful content. In such cases, we repeatedly evaluate the sample, although a few cases may still consistently fail under certain conditions.

We evaluate generated audio from two complementary perspectives. First, for \textbf{Event Coverage}, the evaluator checks each annotated audio event independently and returns a binary judgment indicating whether the event is audible or whether an acoustically equivalent event is clearly present in the generated audio. The final event score is computed as the average pass rate over all annotated events. Second, for \textbf{Temporal Consistency}, the evaluator checks each temporal statement independently. These statements describe relations such as before, after, or overlap between events, and the evaluator determines whether each relation is satisfied in the generated audio. The final temporal score is computed as the average pass rate over all temporal statements.

In addition, we conduct a LALM-based evaluation. The evaluator assigns scores on a 1--5 scale along three dimensions: \textbf{Quality}, measuring audio fidelity, clarity, artifacts, and listening comfort; \textbf{Alignment}, measuring semantic, temporal, spatial, and scene-level consistency with the input description; and \textbf{Aesthetic}, measuring artistic appeal, naturalness of sound design, emotional impact, and pleasantness. The prompts used for evaluation are shown below.

\newtcblisting{promptbox}[2][]{
  enhanced,
  breakable,
  listing only,
  colback=white,
  colframe=black!45,
  colbacktitle=black!7,
  coltitle=black,
  title={#2},
  fonttitle=\bfseries,
  boxrule=0.4pt,
  arc=1mm,
  left=5pt,
  right=5pt,
  top=5pt,
  bottom=5pt,
  listing options={
    basicstyle=\ttfamily,
    breaklines=true,
    breakatwhitespace=true,
    breakautoindent=false,
    breakindent=0pt,
    columns=fullflexible,
    keepspaces=true
  },
  #1
}

\begin{promptbox}{Prompt for Audio Events Coverage Evaluation}
You are evaluating a generated audio sample for a Text-to-Audio (T2A) task.
You will receive the target audio caption and a numbered list of required audio events. And you will receive the generated audio.
Your job is to listen to the generated audio and judge whether each required audio event is present in the generated audio. Use the audio caption only as context for what the T2A system was asked to create. Do not mark an event as passed just because it appears in the caption or event list.
Judging rules:
- passed=true means the event is audible in the generated audio, or an acoustically equivalent event is clearly present.
- passed=false means the event is missing, replaced by a meaningfully different sound, or cannot be reasonably confirmed.
- Keep each reason short and grounded in what you heard.
- confidence must be one of: high, medium, low.
Return strict JSON only. Do not use Markdown or code fences. The JSON schema is:
{
  "items": [
    {
      "index": 1,
      "event": "event text copied from the input",
      "passed": true,
      "reason": "short reason",
      "confidence": "high"
    }
  ]
}
audio_caption:
{audio_caption}
audio_events:
{audio_events}
\end{promptbox}

\begin{promptbox}{Prompt for Temporal Statements Consistency Evaluation}
You are evaluating a generated audio sample for a Text-to-Audio (T2A) task.

You will receive the target audio caption and a numbered list of temporal or relation statements. And you will receive the generated audio.

Your job is to listen to the generated audio and judge whether each statement is satisfied by the generated audio. Use the audio caption only as context for what the T2A system was asked to create. Do not mark a statement as passed just because it appears plausible from the caption.

Judging rules:
- passed=true means the generated audio clearly satisfies the statement.
- passed=false means the generated audio does not satisfy the statement, the relevant sounds are missing, or the statement cannot be reasonably confirmed.
- Judge each statement independently.
- Keep each reason grounded in what you heard.
- confidence must be one of: high, medium, low.

Return strict JSON only. Do not use Markdown or code fences. The JSON schema is:
{
  "items": [
    {
      "id": "T1",
      "statement": "statement text copied from the input",
      "passed": true,
      "reason": "short reason",
      "confidence": "high"
    }
  ]
}

audio_caption:
{audio_caption}

statements:
{statements}
\end{promptbox}

\begin{promptbox}{Prompt for LALM-based Score Evaluation}
You are evaluating a generated audio sample for a subjective Text-to-Audio MOS study.

You will receive the target story/audio caption and the generated audio.

Please listen to the generated audio and score these three dimensions on a 1-5 scale:
- Quality: audio fidelity, clarity, noise level, distortion/artifacts, intelligibility, and listening comfort.
- Alignment: semantic, temporal, spatial, and scene-coherence consistency between the generated audio and the caption/input scene.
- Aesthetic: overall artistic appeal, naturalness of the sound design, emotional impact, and pleasantness.

Scoring scale:
1 = very poor
2 = poor
3 = fair
4 = good
5 = excellent

Use the caption as the target content. Judge what is actually audible in the generated audio. Keep reasons grounded in what you heard.

Return strict JSON only. Do not use Markdown or code fences. The JSON schema is:
{
  "quality": {
    "reason": "reason",
    "score": 1
  },
  "alignment": {
    "reason": "reason",
    "score": 1
  },
  "aesthetic": {
    "reason": "reason",
    "score": 1
  }
}

caption:
{caption}
\end{promptbox}

\end{document}